# Journal of Geophysical Research: Space Physics

## Severe and localized GNSS scintillation at the poleward edge of the nightside auroral oval during intense substorm aurora

Christer van der Meeren[1], Kjellmar Oksavik[1,2], Dag A. Lorentzen[3,4], Michael T. Rietveld[5], and Lasse B. N. Clausen[6]

[1]Birkeland Centre for Space Science, Department of Physics and Technology, University of Bergen, Bergen, Norway, [2]University Centre in Svalbard, Longyearbyen, Norway, [3]Birkeland Centre for Space Science, University Centre in Svalbard, Longyearbyen, Norway, [4]Now at British Antarctic Survey, Cambridge, UK, [5]European Incoherent Scatter Association, Ramfjordbotn, Norway, [6]Department of Physics, University of Oslo, Oslo, Norway

**Abstract** In this paper we study how GPS, GLONASS, and Galileo navigation signals are compromised by strong irregularities causing severe phase scintillation ($\sigma_\phi > 1$) in the nightside high-latitude ionosphere during a substorm on 3 November 2013. Substorm onset and a later intensification coincided with polar cap patches entering the auroral oval to become auroral blobs. Using Global Navigation Satellite Systems (GNSS) receivers and optical data, we show severe scintillation driven by intense auroral emissions in the line of sight between the receiver and the satellites. During substorm expansion, the area of scintillation followed the intense poleward edge of the auroral oval. The intense auroral emissions were colocated with polar cap patches (blobs). The patches did not contain strong irregularities, neither before entering the auroral oval nor after the aurora had faded. Signals from all three GNSS constellations were similarly affected by the irregularities. Furthermore, two receivers spaced around 120km apart reported highly different scintillation impacts, with strong scintillation on half of the satellites in one receiver and no scintillation in the other. This shows that areas of severe irregularities in the nightside ionosphere can be highly localized. Amplitude scintillations were low throughout the entire interval.

## 1. Introduction

Ionospheric scintillations are disturbances on transionospheric communication links such as Global Navigation Satellite Systems (GNSS) signals. Scintillations are caused by irregularities in the plasma density. In the high-latitude ionosphere, such irregularities are known to be associated with auroral particle precipitation, polar cap patches, and auroral blobs [e.g., *Moen et al.*, 2013; *Jakowski et al.*, 2012; *Aarons et al.*, 2000; *Hosokawa et al.*, 2014; *Buchau et al.*, 1985; *Weber et al.*, 1986; *Jin et al.*, 2014]. The current paper provides a multiinstrument case study of an event with severe scintillation in the nightside ionosphere involving auroral precipitation, patches, and blobs. It will be shown that the scintillation can be highly localized and colocated with substorm auroral precipitation. Since the nightside polar cap and auroral ionosphere is a highly dynamic and complicated environment, a brief overview of auroral precipitation, patches, and blobs and their relation to GNSS scintillation is provided next as a background for the case study.

### 1.1. Auroral Emissions and Substorm Activity

An important and well-known feature of the active magnetosphere-ionosphere system is the magnetospheric substorm. A magnetospheric substorm is a transient process initiated on the nightside of the Earth, in which a significant amount of energy from the solar wind-magnetosphere interaction is deposited in the auroral ionosphere and magnetosphere [*Rostoker et al.*, 1980]. This is manifested most visibly as auroral emissions, primarily 557.7nm emissions from $O(^1S)$ at ~120 km altitude and 630.0nm emissions from $O(^1D)$ at ~200–250km altitude [e.g., *Solomon et al.*, 1988]. The substorm is normally divided into three phases called growth, expansion, and recovery [e.g., *McPherron*, 1979, 1970; *Akasofu*, 1964]. A review of the substorm and its phases is provided by *Elphinstone et al.* [1996]. We will highlight points relevant to our study: In the growth phase, energy is loaded into the magnetosphere by a southward oriented interplanetary magnetic field (IMF) and dayside reconnection. The polar cap expands due to the added open flux, and the auroral oval migrates equatorward. In the expansion phase, energy stored in the magnetotail is explosively released into the ionosphere. The aurora suddenly brightens and expands poleward as the magnetotail performs a dipolarization. The energetic particle precipitation enhances the conductivity in the ionosphere, which causes a sudden enhancement of the eastward and westward auroral electrojets. This enhancement is detectable in ground magnetometers due to the westward/eastward electrojet causing a negative/positive deviation in the horizontal component of the magnetic field above the station. The third phase of the substorm is the recovery phase, during which the intensity of the auroral emissions are reduced. The duration of the entire substorm cycle is on the order of a few hours, though during periods of continuous southward IMF the recovery phase may coincide with the growth phase of the next substorm.

### 1.2. Polar Cap Patch Activity

Another well-known high-latitude feature during active geomagnetic conditions is polar cap patches [e.g., *Carlson*, 2012, and references therein]. During periods of southward IMF, a two-cell convection pattern is set up in the ionosphere which can convect plasma from the solar-ionized high-density plasma reservoir in the dayside ionosphere, through the cusp region, and across the polar cap to the nightside auroral oval [*Dungey*, 1961; *Weber et al.*, 1984; *Foster and Doupnik*, 1984; *Buchau et al.*, 1985; *Foster*, 1993; *Foster et al.*, 2005; *Moen et al.*, 2008; *Cousins and Shepherd*, 2010; *Oksavik et al.*, 2010; *Zhang et al.*, 2013a; *Nishimura et al.*, 2014; *van der Meeren et al.*, 2014]. This plasma is frequently segmented upon entry to the polar cap, with magnetopause reconnection proposed as the dominant segmentation mechanism [*Lockwood and Carlson*, 1992; *Carlson et al.*, 2002, 2004, 2006; *Lockwood et al.*, 2005; *Moen et al.*, 2006; *Lorentzen et al.*, 2010; *Zhang et al.*, 2013b]. The resulting islands of enhanced plasma density are called F region polar cap patches. These are typically 100–1000km across and show up as regions of 630.0nm emissions from dissociative recombination of $O_2^+$ with F region electrons creating $O(^1D)$ [e.g., *Wickwar et al.*, 1974; *Hosokawa et al.*, 2011]. These airglow emissions are frequently detectable from ground-based optical instruments [e.g., *Buchau et al.*, 1983; *Weber et al.*, 1984; *Lorentzen et al.*, 2004; *Hosokawa et al.*, 2006; *Moen et al.*, 2007; *Nishimura et al.*, 2014; *van der Meeren et al.*, 2014]. In the nightside ionosphere, patches convect into the nightside auroral oval under the influence of tail reconnection. Patches convecting into the auroral oval have been shown to be associated with substorm onset [*Lyons et al.*, 2011; *Nishimura et al.*, 2013, 2014; *Shi and Zesta*, 2014]. In the auroral oval, they are termed auroral blobs [*Tsunoda*, 1988; *Crowley et al.*, 2000; *Lorentzen et al.*, 2004; *Pryse et al.*, 2006]. Several types of auroral blobs are referred to in literature, namely, boundary blobs, subauroral blobs, and auroral blobs [e.g., *Crowley et al.*, 2000]. In this study, following the example of *Jin et al.* [2014], we use the term to describe patches inside the active auroral oval.

### 1.3. Ionospheric Scintillations

Both patches and auroral emissions are known to be associated with disturbances of transionospheric signals, also termed scintillations (these relationships will be detailed shortly). Scintillations are rapid variations in the amplitude or phase of radio signals, such as global navigation satellite system (GNSS) signals, and are associated with decameter- to kilometer-scale irregularities in the ionosphere [e.g., *Hey et al.*, 1946; *Basu et al.*, 1990, 1998; *Kintner et al.*, 2007]. Scintillations are categorized into amplitude and phase scintillations. Amplitude scintillations are caused by irregularities with scale sizes of tens of meters to hundreds of meters, more precisely at and below the Fresnel radius, which is approximately 360m for GPS L1 frequency (1575.42MHz) and an irregularity altitude of 350km [e.g., *Forte and Radicella*, 2002]. Amplitude scintillations are normally quantified by the dimensionless $S_4$ index, which is the standard deviation of the received power $I$ normalized by its mean value over some period of time [*Briggs and Parkin*, 1963]:

$$S_4^2 = \frac{\langle I^2 \rangle - \langle I \rangle^2}{\langle I \rangle^2}$$

Phase scintillations are caused by irregularities with scale sizes from a few hundred meters up to several kilometers [e.g., *Kintner et al.*, 2007] and are normally quantified by the $\sigma_\phi$ index, which is the standard deviation of the detrended carrier phase $\phi$ in radians over some period of time [*Fremouw et al.*, 1978]:

$$\sigma_\phi^2 = \langle \phi^2 \rangle - \langle \phi \rangle^2$$

The period is normally 60s, and the detrending is usually done using a sixth-order Butterworth filter with a cutoff frequency of 0.1Hz [e.g., *Mitchell et al.*, 2005; *Béniguel et al.*, 2009; *Li et al.*, 2010; *Alfonsi et al.*, 2011;

*Forte et al.*, 2011; *Garner et al.*, 2011; *Gwal and Jain*, 2011; *Kinrade et al.*, 2012; *Jiao et al.*, 2013; *Jin et al.*, 2014]. This is not without problems, as the $\sigma_\phi$ index is highly sensitive to the cutoff frequency, and 0.1Hz has been shown to be problematic at high latitudes [e.g., *Forte*, 2005]. However, it is possible to get a better overview of the phase variations at different scales by looking at a spectrogram of the raw phase, as was done by *van der Meeren et al.* [2014]. While phase scintillation from satellites in polar orbits are heavily influenced by geometrical factors (specifically, scintillation levels are significantly enhanced when the signal path is *L* shell aligned), this has been shown not to be important at high latitudes for satellites in GPS-like orbits [*Forte and Radicella*, 2004].

Scintillations are found where ionospheric irregularities occur, predominantly in the equatorial and auroral/ polar regions [*Basu et al.*, 1988]. High-latitude scintillation is correlated with solar cycle [*Basu et al.*, 1988; *Skone*, 2001] and magnetic activity [*Tiwari et al.*, 2012] and occurs primarily in the cusp and nightside auroral oval [*Kersley et al.*, 1995; *Prikryl et al.*, 2011; *Jin et al.*, 2015]. Phase scintillation is more prominent than amplitude scintillation in the polar ionosphere [*Spogli et al.*, 2009; *Prikryl et al.*, 2010; *Gwal and Jain*, 2011].

### 1.4. Scintillation From Auroral Activity

*Aarons et al.* [2000] was the first to show a correlation between fluctuations in GPS phase and substorm-related auroral disturbances. Statistical studies have found that the auroral region is more sensitive to phase than amplitude scintillation and that the phase scintillation occurs at lower latitudes during perturbed conditions, implying a close relationship with the auroral oval [*Spogli et al.*, 2009; *Tiwari et al.*, 2012; *Jiao et al.*, 2013]. Indeed, *Spogli et al.* [2009] found an enhancement of scintillation associated with the boundaries of the statistical auroral oval. It has also been found that auroral scintillation occurs most frequently near midnight magnetic local time (MLT) and that the occurrence and magnitude is strongly correlated with the disturbance of the local magnetic field [*Jiao et al.*, 2013]. *Kinrade et al.* [2013] correlated auroral emissions (557.7nm and 630.0nm) with phase scintillation on a geographical grid and found a proxy relationship between optical emissions and $\sigma_\phi$ in the presence of strong auroral activity.

Case studies support the statistical findings. Phase scintillation has been associated with auroral arc brightening and substorms [*Prikryl et al.*, 2010; *Ngwira et al.*, 2010]. *Hosokawa et al.* [2014] found that phase scintillation was enhanced in relation to substorm onset and decreased as the aurora became more diffuse. They suggested that discrete aurora in the GPS signal path is necessary for the occurrence of phase scintillation during substorm intervals. In addition to scintillation, other adverse effects such as loss of lock [*Smith et al.*, 2008] and cycle slips [*Prikryl et al.*, 2010] have been directly observed in relation to auroral emissions.

### 1.5. Scintillation From Polar Cap Patches

It has long been known that polar cap patches are source regions for decameter- to kilometer-scale irregularities causing scintillation [*Buchau et al.*, 1985; *Weber et al.*, 1986; *Basu et al.*, 1990, 1991, 1994, 1998; *Coker et al.*, 2004; *Carlson*, 2012, and references therein]. It has been suggested that the patches may be initially structured through the Kelvin-Helmholtz instability in the dayside patch segmentation region, creating seed irregularities which allow the gradient drift instability to effectively structure the patch at smaller scale sizes during its transit across the polar cap [*Carlson et al.*, 2007, 2008]. While the gradient drift instability is most effective on the trailing edge of density structures, simulations have shown that the irregularities may propagate from the trailing edge into the interior of the patch, structuring the whole patch at a variety of scale sizes [*Gondarenko and Guzdar*, 2004]. This is supported by observations [e.g., *Hosokawa et al.*, 2009]. Case studies have found an agreement between scintillation and optical airglow from patches [*Jin et al.*, 2014; *Coker et al.*, 2004].

Statistically, an agreement has been found between scintillations at polar latitudes and the asymmetric distribution of patches around magnetic midnight [*Spogli et al.*, 2009], as well as with the IMF dependence of polar cap patches [*Li et al.*, 2010]. Patches are linked to both phase and amplitude scintillation [*Alfonsi et al.*, 2011].

### 1.6. Motivation for This Study

It is not yet clearly established which ionospheric phenomena produce the strongest irregularities associated with problematic scintillation. *Jin et al.* [2014] studied scintillation from polar cap patches, auroral arcs, and auroral blobs (patches in the auroral oval). They concluded that the most severe scintillation in the European Arctic sector are due to blobs, i.e., when patches are structured by substorm auroral arc dynamics. Specifically, they found that auroral blobs were associated with the strongest scintillation, followed by polar cap patches on their own, with auroral arcs alone showing the least scintillation. That was, however, a case study of only a single night, and further studies are needed to ascertain whether their observations are representative.

Furthermore, while there have been many statistical studies of scintillation-producing irregularities, the temporal and spatial averaging inherent to these studies means it is difficult to clearly establish a direct link between scintillations and highly transient phenomena such as polar cap patches and auroral arc dynamics.

A notable space weather impact of intense scintillation is loss of signal lock on GNSS signals, potentially resulting in the receiver being unable to compute a navigation solution if sufficiently many signals are affected [e.g., *Kintner et al.*, 2007]. The propensity for losing lock is receiver-dependent and may be influenced by the internal state of the receiver [*Garner et al.*, 2011]. Several GNSS constellations exist, e.g., GPS, GLONASS, and Galileo. For multiconstellation GNSS receivers, this increases the likelihood that a sufficient number of signals will remain trackable during scintillation events. However, the signals of all three constellations use similar frequencies in the L band, and one should therefore expect that all constellations are prone to similar scintillation effects.

This case study uses three all-sky imagers and 50Hz raw data from four multiconstellation GNSS receivers in the Svalbard region to provide a detailed look at when and where scintillation occurs during substorm activity in the nightside polar cap. The study shows that severe scintillation is observed when auroral precipitation coincides with patches. Our observations further indicate that the irregularities are driven by auroral precipitation and that the scintillation is highly localized. This is also the first study to show directly that GPS, GLONASS, and Galileo are similarly affected by severe scintillation in relation to intense line-of-sight auroral emissions in a highly localized region of the sky.

## 2. Instrumentation

### 2.1. GNSS Receivers

The GNSS data used in this study come from four NovAtel GPStation-6 GNSS Ionospheric Scintillation and TEC Monitors installed in Svalbard in 2013 and operated by the University of Bergen. The receiver locations, geographic latitudes (GLAT), geographic longitudes (GLON), and magnetic latitudes (MLAT, Altitude Adjusted Corrected Geomagnetic Coordinates (AACGM) [*Baker and Wing*, 1989]) are Ny-Ålesund (78.9° GLAT, 11.9° GLON, 76.4° MLAT), Longyearbyen (78.1° GLAT, 16.0° GLON, 75.4° MLAT), Hopen (76.5° GLAT, 25.0° GLON, 73.3° MLAT), and Bjørnøya (74.5° GLAT, 19.0° GLON, 71.6° MLAT).

Magnetic midnight is around 2100–2130 UT. All receivers track GPS, GLONASS, and Galileo at several frequencies. For the period of study, all constellations were tracked at L1 (1575.42MHz). Additionally, GLONASS was tracked at L2 (1227.60MHz) and Galileo was tracked at E5 (1191.795MHz). There was no significant difference in scintillation levels in the different signals (the lower frequencies showed marginally higher levels), so L1 will be used in this study.

The receivers output both 60s reduced data and 50Hz raw data (phase and power). In this paper, both are used. Specifically, 50Hz raw data are used to compute the $\sigma_\phi$ index with 1s resolution. When calculating the phase scintillation index, the raw phase is detrended by subtracting a polynomial fit and filtered using a sixth-order Butterworth high-pass filter with a cutoff frequency of 0.1Hz. The $S_4$ index is calculated by subtracting $S_4$ due to ambient noise from the total $S_4$ in a root-sum-square sense according to the user manual [*NovAtel*, 2012]. A minimum lock time of 60s was required for $S_4$ data, and 240s for $\sigma_\phi$. The elevation was required to be at least 20° to limit multipath effects.

The GNSS ionospheric piercing points (IPPs) have been projected to 150km altitude when shown together with green 557.7nm aurora and 250km when shown together with red 630.0nm emissions.

In the same manner as *van der Meeren et al.* [2014] and *Oksavik et al.* [2015], spectrograms of the raw phase are used in order to obtain more detailed information on the phase variations. The spectrograms were made using wavelet analysis, based on software provided by *Torrence and Compo* [1998]. The Morlet wavelet was chosen as the mother wavelet. This method has been used previously by other GNSS studies [e.g., *Mushini et al.*, 2012]. No detrending of the raw phase is required to produce the wavelet spectrograms. For the reader's information, the wavelet spectrograms were compared to spectrograms made using Fourier analysis of detrended data and showed exactly the same features. The wavelet spectrograms provided much better resolution at smaller scales and were thus chosen for this study. For further details on wavelet analysis we refer to the literature [e.g., *Torrence and Compo*, 1998; *Mushini et al.*, 2012].

### 2.2. Optics

We use data from three all-sky imagers (ASI) and a meridian-scanning photometer (MSP). The fields of view at 250km altitude are indicated in red in Figure 1.

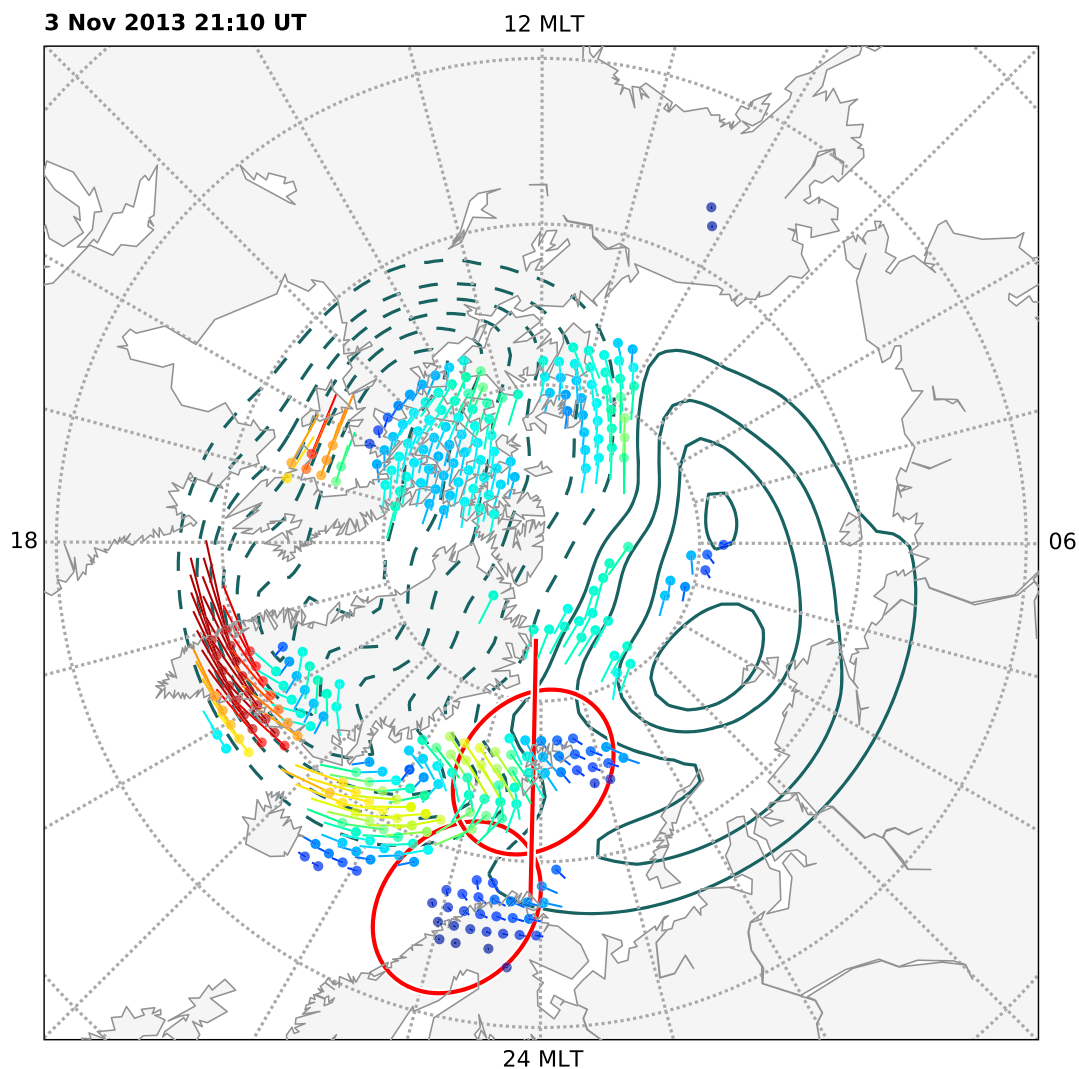

**Figure 1.** SuperDARN drift vectors and polar cap potential data at 2110 UT showing a two-cell convection pattern (representative of the whole period). The data are displayed in an MLAT/MLT grid. The red line and circles indicate the fields of view of the meridian-scanning photometer in Longyearbyen and the all-sky imagers in Longyearbyen and at Andøya projected to 250 km altitude.

The ASIs are operated by the University of Oslo and are located in Ny-Ålesund and Longyearbyen (colocated with the GNSS receivers) and at Andøya (69.2° GLAT, 16.0° GLON, 66.3° MLAT). All three imagers have filters for green line (557.7 nm) and red line (630.0 nm) emissions. The imager in Longyearbyen is calibrated. Emissions at 557.7 nm are projected to 150 km altitude, while emissions at 630.0 nm are projected to 250 km.

The MSP is located in Longyearbyen and is operated by the University Centre in Svalbard. It scans along the magnetic meridian and records a full scan at a 16 s cadence of green (557.7 nm) and red (630.0 nm) emissions. The intensity is calibrated.

### 2.3. Solar Wind and the *AE* Index
Data from the IMF and solar wind, as well as the auroral electrojet (*AE*) index, are provided by the NASA OMNI-Web service. Both the IMF and plasma data are provided by the Wind spacecraft [*Lepping et al.*, 1995; *Ogilvie et al.*, 1995]. The spacecraft was located at $(X, Y, Z) = (254, 48, 20) R_E$ (geocentric solar ecliptic coordinates). The data are time shifted to the bow shock by the OMNIWeb service.

### 2.4. Magnetometer Data
We make use of the horizontal component of the magnetic field measured at Bjørnøya (BJN). The magnetometer is a fluxgate magnetometer colocated with the GNSS receiver at Bjørnøya.

### 2.5. SuperDARN and Convection Data
The Super Dual Auroral Radar Network (SuperDARN) is a network of coherent high-frequency (HF) scatter radars measuring backscatter from field-aligned decameter-scale irregularities [*Greenwald et al.*, 1995; *Chisham et al.*, 2007]. HF backscatter is therefore an indicator of the presence of decameter-scale irregularities. Drift velocity data are used to calculate the polar cap potential and convection pattern. The SuperDARN data is retrieved from Virginia Tech using the DaViTpy software package.

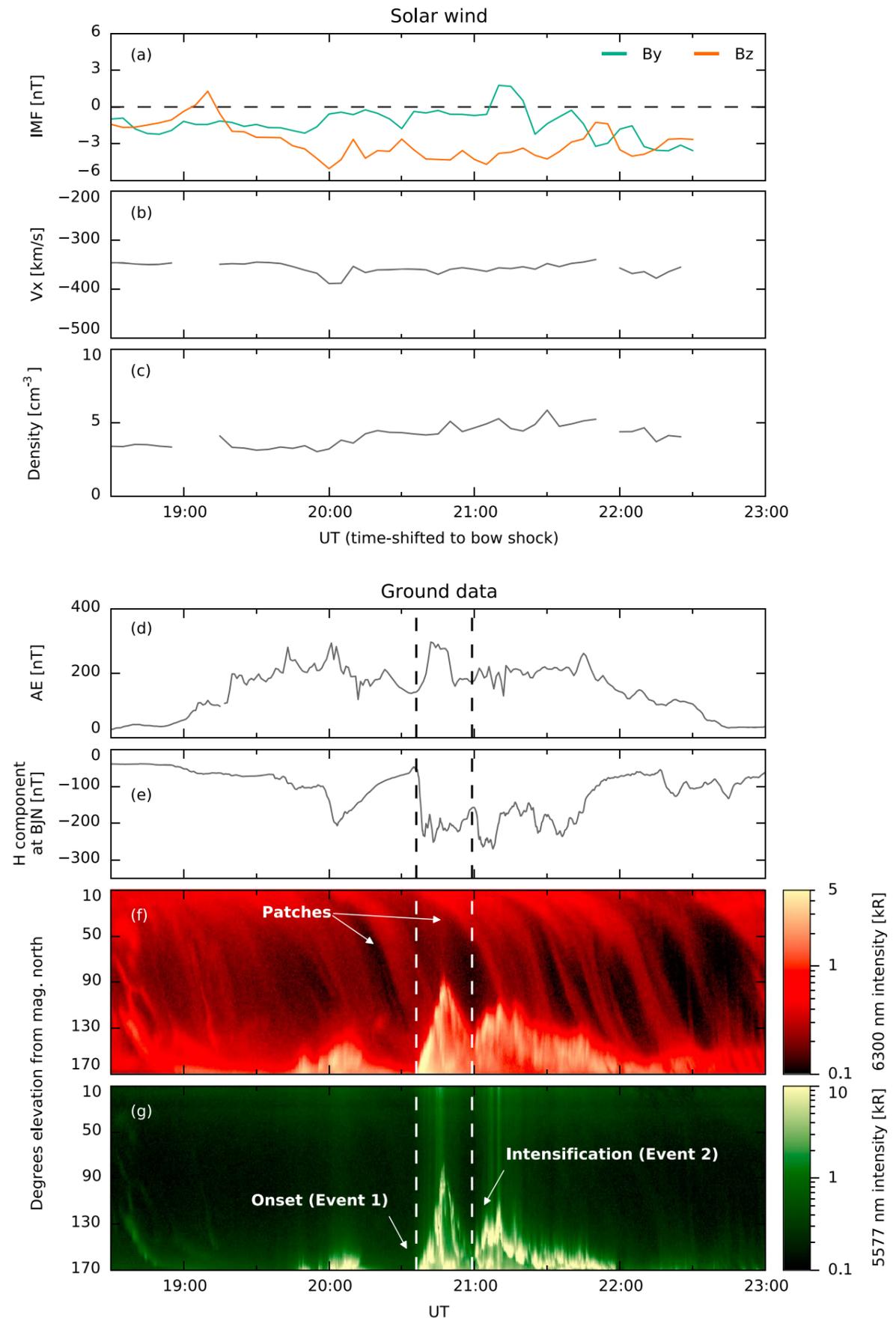

**Figure 2.** Overview of the geomagnetic conditions. (a) The $B_y$ and $B_z$ GSM (geocentric solar magnetic) components of the interplanetary magnetic field, (b) the radial solar wind speed, and (c) the solar wind density. The data are from the Wind satellite at $(X, Y, Z) = (254, 48, 20) R_E$ and have been time shifted to the bow shock by OMNIWeb. (d) The $AE$ (auroral electrojet) index. (e) The deviation in the horizontal $H$ component of the geomagnetic field at Bjørnøya (BJN). The (f) 630.0 nm and (g) 557.7 nm emissions as observed by the meridian-scanning photometer in Longyearbyen between elevation angles of 10° (magnetic north) and 170° (magnetic south). The vertical lines mark the substorm onset and intensification.

## 3. Observations

Figure 2 shows an overview of the solar wind and geomagnetic conditions on 3 November 2013 between 1830 and 2300 UT. Figure 2a shows the $B_y$ and $B_z$ components of the IMF. The IMF was southward oriented, which is known to be associated with a two-cell convection pattern across the polar cap. This is exemplified by Figure 1, which shows the polar cap potential from the SuperDARN radars at 2110 UT. Figures 2b and 2c indicate stable conditions in the solar wind speed and density, respectively.

The *AE* index (Figure 2d) shows increased magnetic activity during a period of around 4 h. The *Kp* index was around 2 (not shown), which indicates weakly disturbed geomagnetic conditions. Figure 2e shows the deviation of the horizontal (*H*) component of the magnetic field at Bjørnøya (BJN). The *H* component drops sharply around 2035 UT, which indicates a sudden enhancement of the westward electrojet over this station consistent with ongoing substorm activity. Another sharp drop is seen around 2100 UT, coinciding with a new intensification of the aurora.

The MSP data (Figures 2f and 2g) clearly show the poleward expansion of the aurora after onset, as well as an intensification and a slight expansion around 2100 UT. In this study, we will refer to the onset and expansion as Event 1, and the intensification as Event 2. The red line emissions show several patches drifting southward from the polar cap and into the auroral oval. Two of these patches (indicated in the figure) coincide with the substorm onset and the intensification, respectively. The MSP data also show a brightening around 2000 UT, which may be a pseudo-onset. While this was accompanied by fairly strong scintillation ($\sigma_\phi$ up to 0.7–0.8, not shown), the observation geometry was not favorable for this event due to the aurora being too low on the horizon. In this study we will only consider the aforementioned onset and intensification events.

Figure 3 gives a detailed view of auroral intensity and phase scintillation at four different times during Event 1 (onset/expansion). Figures 3a–3d and 3e–3h show red line and green line emissions, respectively. Figures 3a and 3e are before onset and shows that $\sigma_\phi < 0.2$ for all satellites and all four receivers. This includes the patch, which can be seen in the red line emissions (Figure 2a) as a light blue area stretching from west to northeast of Svalbard. According to ionosonde data from Longyearbyen (not shown), the patch has a critical frequency of 6.2MHz (corresponding to an electron density of $4.8 \times 10^{11}$ m$^{-3}$), which is 1.5 times the background density. (For the reader's information, total electron content (TEC) data were studied but were inconclusive both in the patch and auroral regions and were not found to be helpful to the current study.) Figures 3b–3c and 3f–3g show scintillation during the poleward expansion of the aurora. The patch drifts southward, and as the southwestern edge of the patch reaches the auroral oval, onset and poleward expansion of the aurora occurs. This can be seen in Figure 2b but is more easily seen in the supporting information Movie S1. The phase scintillation follows the intense poleward edge of the aurora as it expands northward (across the patch) and reaches values of $\sigma_\phi > 1$ during the expansion. There are only minor scintillations in the more diffuse emissions further equatorward of the edge. Finally, the rightmost column shows scintillation after the expansion has subsided and the aurora has faded. Due to the auroral emissions, it is not possible to see the patch in the optical data. There is no significant scintillation ($\sigma_\phi < 0.2$ for almost all satellites).

Figure 4 is in the same format as Figure 3 and shows the auroral intensity and phase scintillation during Event 2 (the intensification). Figures 4a and 4e show the conditions before the intensification. A patch is seen as a light blue area to the northwest of Svalbard. The peak density of the patch is difficult to determine due to only a small part of it drifting over the ionosonde in Longyearbyen, but a later part of the patch drifting over Longyearbyen at 2122 UT was studied and showed a critical frequency of 6.0MHz (electron density $4.5 \times 10^{11}$ m$^{-3}$) over a background of 4.0MHz ($2.0 \times 10^{11}$ m$^{-3}$), which is a relative density of 2.3. There is no scintillation ($\sigma_\phi < 0.2$) in relation to the patch or the dim aurora. The exact time of arrival of the patch at the poleward edge of the aurora is difficult to ascertain due to the patch entering the aurora at an oblique angle, but the arrival roughly coincides with an auroral intensification and strong scintillations. The brightest aurora and strongest scintillation occurs around 2110 UT (column 3). Here the aurora peaks at 55kR, and $\sigma_\phi > 1.5$ in the most intense aurora for satellites from all three constellations. As in the previous event, the patch is colocated with the strong aurora and the scintillations. Figures 4d and 4h show that there are no strong scintillations after the aurora has faded and the patch has entered the auroral zone.

Data from a GPS receiver at Hornsund (not shown) have also been studied and support these findings. For the reader's reference, an animation in the style of Figures 3 and 4 for the period 2000–2130 UT is provided in the supporting information Movie S1.

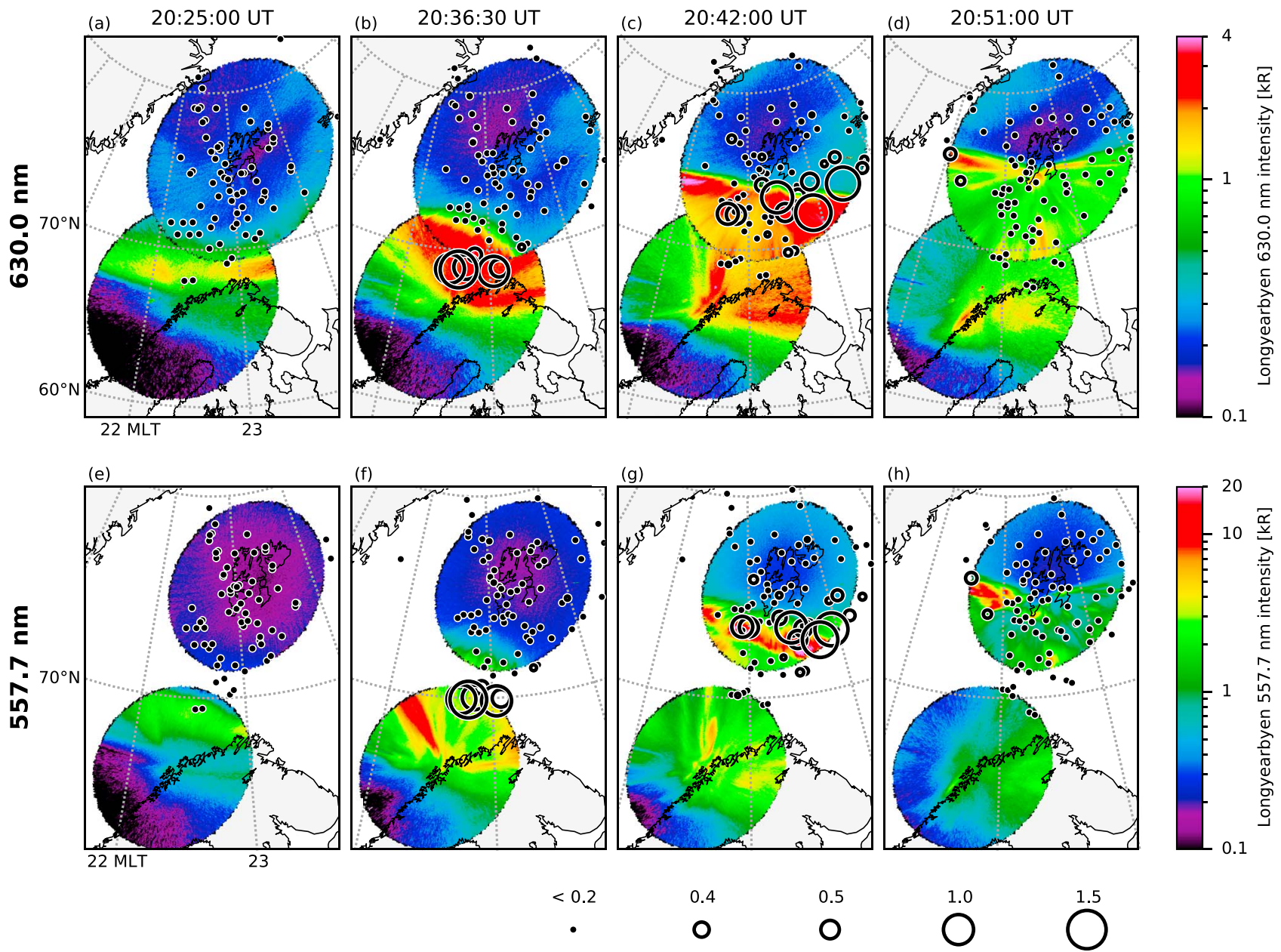

**Figure 3.** Auroral emissions and phase scintillation (a, e) before, (b–c and f–g) during, and (d, h) after Event 1 (the substorm expansion). Panels (a–d) and (e–h) show red line (630.0nm) and green line (557.7nm) emissions, respectively. The GNSS ionospheric piercing points are shown as circles, with the size representing the intensity of the phase scintillation. The arriving patch can be seen in Figure 3a as a light blue area stretching from west to northeast of Svalbard.

Figures 5 and 6 provide a closer look at scintillation and phase variations in relation to line-of-sight auroral emissions for three selected satellites at two receiver locations. In these figures we have computed high-resolution scintillation indices over periods of 1s from the 50Hz raw data. These high-resolution data match closely the corresponding lower resolution 60s data (not shown) and provide a more fine-grained view of the observations. In Longyearbyen (Figure 5), there is little or no amplitude scintillation. However, strong phase scintillation is observed in relation to Event 1, and severe phase scintillation ($\sigma_\phi > 1.5$) is observed in relation to Event 2. The most intense line-of-sight emissions are seen in Event 2. The intensity in Figures 5a–5c (and Figures 6a–6c) is based on a 7-by-7 pixel window centered at the elevation and azimuth of the satellite, and the panels show the median of the pixels as lines and the minimum and maximum of the window as shaded regions. The data show a close correspondence between intense 557.7nm emissions and phase scintillation. In Figure 5m, the severe scintillation is observed for a cluster of five satellites including GPS, GLONASS, and Galileo, three of which are shown in Figures 5a–5l. These signals are intersecting a region of intense 557.7nm emissions (up to 55kR). Strong scintillations are also seen in another satellite in the aurora further east. The phase spectra show enhanced phase variations at a variety of temporal scales (1–50s) during the two events, with short bursts of highly localized variations at smaller scales concurrent with the most intense phase scintillation and auroral emissions.

Figure 6 shows the same type of data from Ny-Ålesund. Since Ny-Ålesund is north of Longyearbyen, the aurora in the south is observed at a lower elevation. Again, there is no amplitude scintillation. The southernmost of the three satellites (GPS 09) show severe phase scintillation ($\sigma_\phi \sim 1$) in relation to Event 1, when the aurora is at its most poleward point during the whole period under study (cf. the MSP data in Figures 2f and 2g). Around this time, the auroral emissions saturate the ASI (Figure 6c). However, unlike in Longyearbyen, only very weak phase scintillation is produced during Event 2. From Figure 6m we observe that the intense 557.7nm emissions during the intensification do not cross the line of sight between the receiver and the satellites. The spectra show enhanced variations at a range of frequencies, but the variations are less severe (lower spectral power), less temporally localized, and less spectrally wide than in the Longyearbyen data.

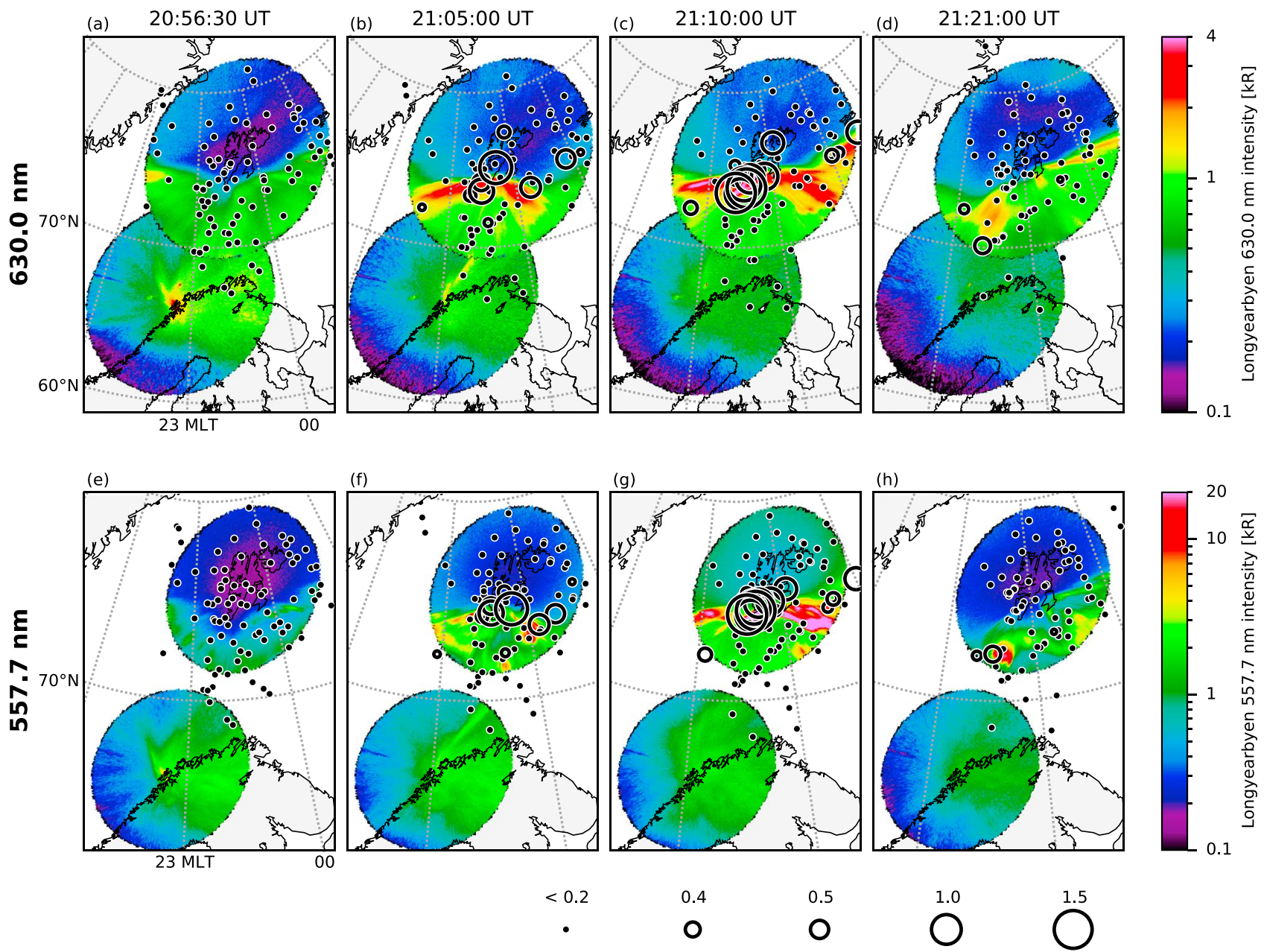

**Figure 4.** Auroral emissions and phase scintillation before, during, and after Event 2 (the intensification). See Figure 3 for description.

None of the four receivers experienced loss of lock above 25° elevation during the whole period (2030–2130 UT).

## 4. Discussion

### 4.1. Comparison of Aurora, Patches, and Auroral Blobs

The severe scintillation in this study appears to be driven by intense auroral emissions. For both events, Figures 5a–5i shows a remarkable correspondence between phase scintillations and the intensity of line-of-sight auroral 557.7nm emissions, as suggested by *Hosokawa et al.* [2014]. For Event 1, Figure 3 shows clearly that the severe phase scintillation follows the intense poleward edge of the auroral oval. This implies a close relation between the scintillations and the poleward expanding arc of auroral emissions. This is in line with previous statistical studies, which have shown a relationship between scintillation and the auroral oval in general [*Tiwari et al.*, 2012; *Jiao et al.*, 2013; *Kinrade et al.*, 2013] and the auroral oval walls in particular [*Spogli et al.*, 2009]. Case studies have also indicated a relationship between auroral emissions and GNSS scintillations [*Prikryl et al.*, 2010; *Ngwira et al.*, 2010; *Kinrade et al.*, 2013]. For our events, Figures 3 and 4 show that there is no scintillation in relation to the patches. Together with Figures 5m and 6m they support the conclusion that the scintillation is driven by the aurora and not the patches: In Ny-Ålesund (Figure 6m), the cluster of five satellites to the southwest do not intersect the aurora, and there is no scintillation. In Longyearbyen (Figure 5m), the signals do intersect the aurora and there is severe scintillation. The patch (not visible in these 557.7nm images) is intersected by the signals at both locations (it arrives first at Ny-Ålesund and then in Longyearbyen). It is thus clear from our observations in the nightside polar cap that the observed patches on their own are not sufficient for creating scintillation-inducing irregularities. This suggests that the auroral precipitation is providing the energy input for the irregularities.

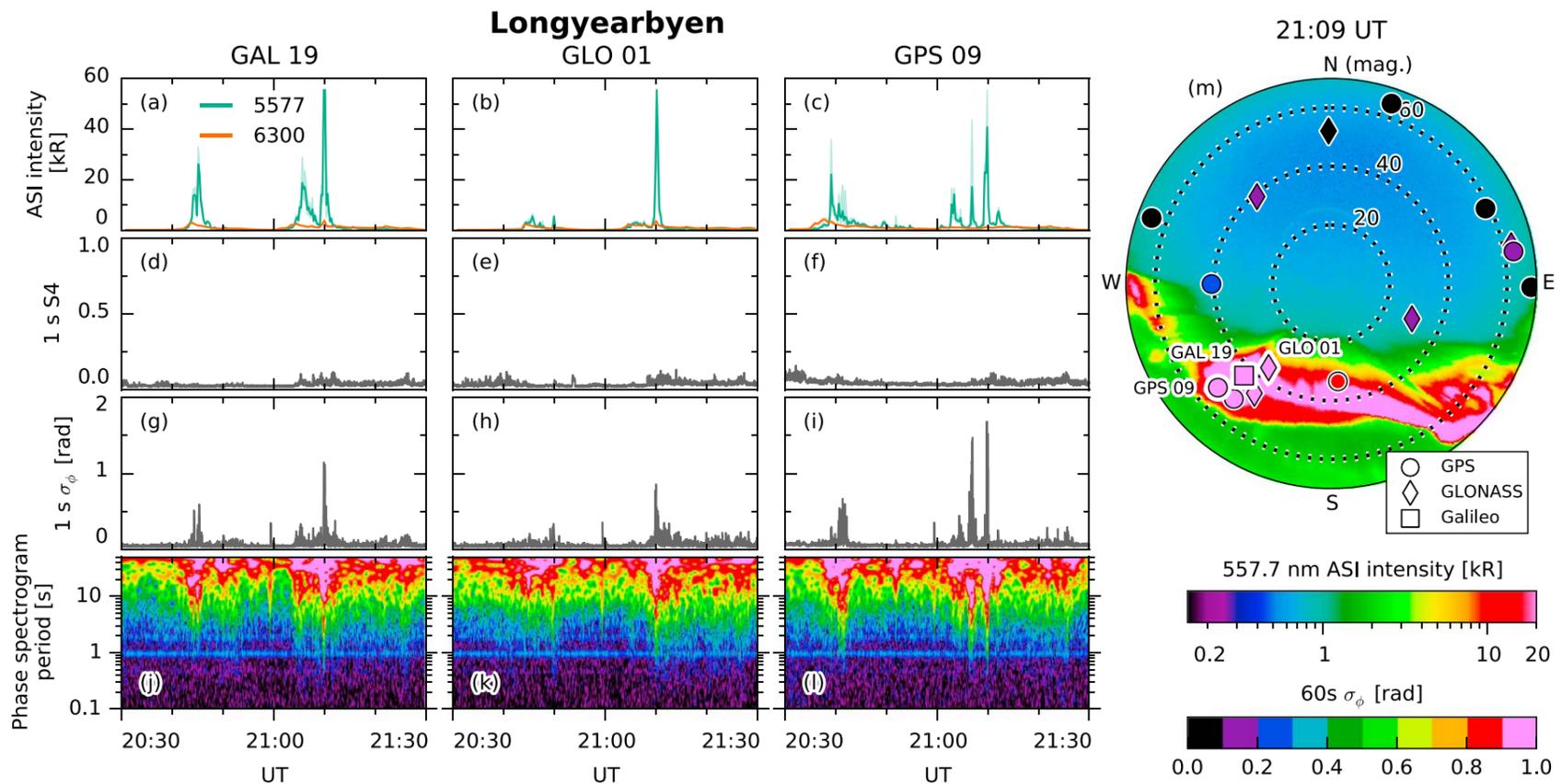

**Figure 5.** GNSS scintillations and phase variations for three selected satellites in relation to line-of-sight auroral intensity at Longyearbyen. (a–c) The auroral intensity in the vicinity of the satellite IPPs. (d–f) The $S_4$ amplitude scintillation index computed from 50Hz raw data over periods of 1s. (g–i) The $\sigma_\phi$ phase scintillation index computed from 50Hz raw data over periods of 1s. (j–l) Wavelet power spectra of 50Hz raw phase on a decibel (logarithmic) color scale (not shown). (m) GNSS satellites and a selected 557.7nm all-sky image on a polar axis (zenith angle versus azimuth; magnetic north is up, magnetic east is right). The three satellites shown in Figures 5a–5l are highlighted.

While the presence of the scintillation can be mainly attributed to the precipitation, the magnitude of the scintillation is still an outstanding question in these events. Specifically, is the auroral precipitation sufficient to produce the observed severe scintillation, or do the blobs (patches in the auroral region) contribute to the strength of the irregularities and scintillation? Fully answering this question requires further studies and more events with both aurora-only and aurora/blob events. To address this issue in a cursory manner, we can compare our results with those of *Jin et al.* [2014]. One of the events they studied was a period of scintillation from aurora without patches. Calibrated ASI data were not available when that study was performed, but the ASI in Longyearbyen has since been calibrated. A brief analysis of line-of-sight intensities and corresponding scintillation (using the Longyearbyen GNSS receiver in the current study) in the manner of Figure 5 indicated values up to 50 kR with corresponding $\sigma_\phi$ levels around 0.2–0.3 (not shown). The intensity values are similar to those of our events, but the scintillation levels are significantly lower. This may indicate that intense aurora alone is not sufficient to produce the severe scintillation we observe in the current study and that the patches/blobs in our events have contributed to the severe scintillation levels observed in the auroral region. If this is the case, we do not have the data to infer the process behind this contribution. Our data do, however, show that the scintillation quickly disappears when the aurora fades. This suggests that if there is such a process, it appears to be directly driven by the auroral precipitation, and the irregularities created by it quickly dissipate when the aurora fades. Further studies are needed to ascertain whether blobs in the auroral region really do contribute to scintillation levels and, if they do, find a physical process behind such a contribution.

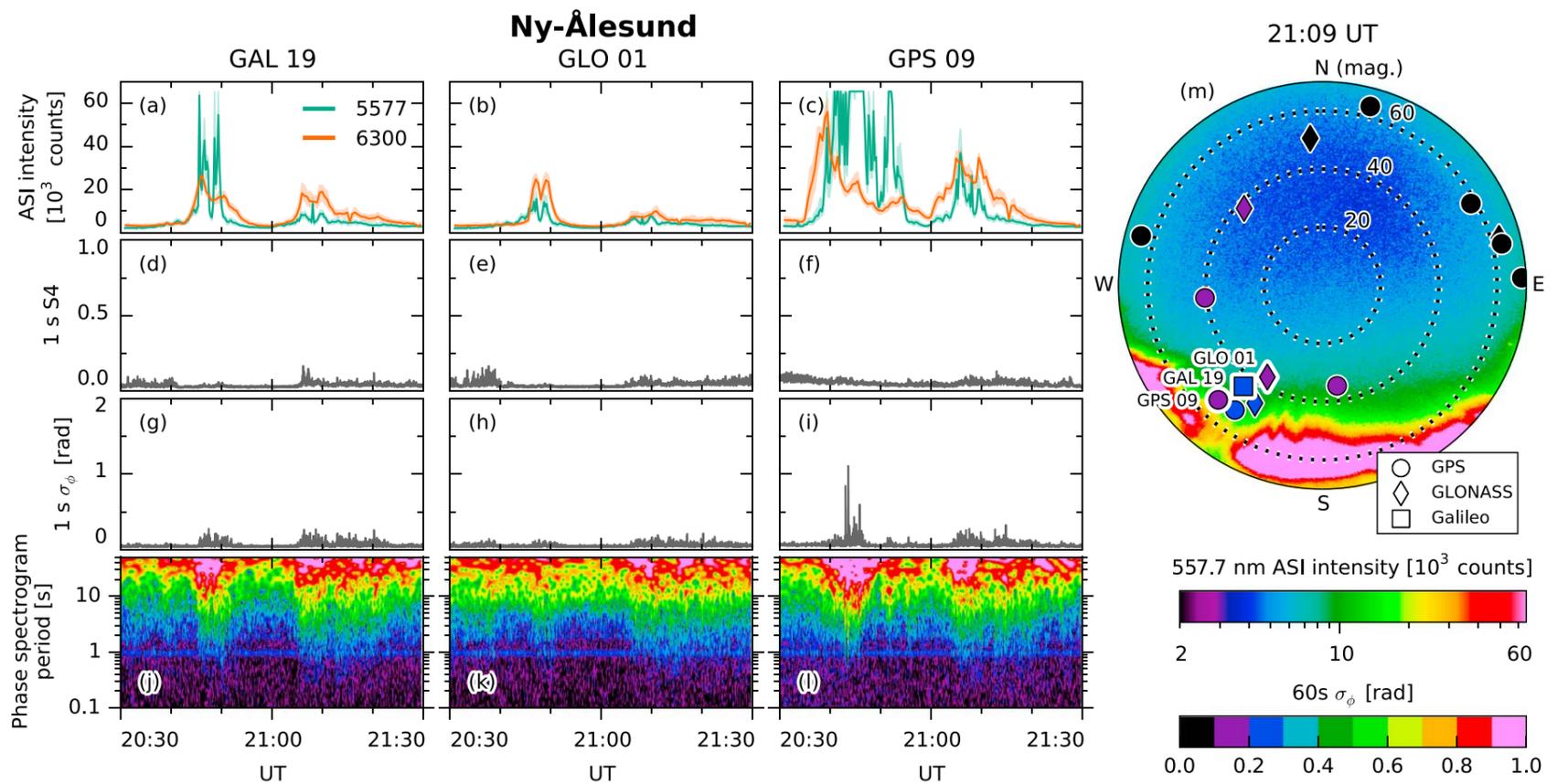

**Figure 6.** Same as Figure 5 but for the Ny-Ålesund receiver and all-sky imager. Note that the emission intensity (Figures 6a–6c) is uncalibrated for both color channels.

The critical reader may be interested to know that while our patches are not associated with scintillation before entering the auroral oval, and the patches studied by *Jin et al.* [2014] are, the patches in the two studies are otherwise similar: Ionosonde data from the period studied by *Jin et al.* [2014] (not shown) are not entirely conclusive due to many *E* region echoes but suggest patch critical frequencies of 6.0–7.2 MHz (corresponding to peak densities of $4.4–6.4 \times 10^{11}$ m$^{-3}$), which is 1.5–2.7 times the respective background densities. This is similar to the absolute and relative densities in our study and was not reported in their paper.

### 4.2. Irregularity Scales

*van der Meeren et al.* [2014] studied a drifting *F* region plasma structure which, due to the low recombination rate in the *F* region, could be viewed as a stable structure (not changing in time) compared to its drift speed across the signal line of sight. Based on the relative drift speed between the plasma and the IPPs, it was therefore possible to convert the temporal scale size of the phase variations to spatial scale and get valuable quantitative information on the scale sizes of the irregularities. Unfortunately, this is not straightforward in this study. The scintillations probably have a strong contribution from energetic particle precipitation in the *E* region. In this region the recombination rate is much higher than in the *F* region, and the plasma cannot be assumed to be a stable, drifting structure. Additionally, there are temporal variations in the auroral precipitation which causes variation in the irregularities. However, since amplitude scintillation is most sensitive to irregularities with scale sizes at and below the Fresnel radius, which does not depend on the drift speed and is on the order of ∼300 m [e.g., *Forte and Radicella*, 2002], the lack of amplitude scintillation implies weaker irregularities at 10–100 m scale. Decameter-scale irregularities may also be studied using HF backscatter data. SuperDARN backscatter from the polar cap was sporadic, and we were not able to see the patches in the radar data. However, the Hankasalmi SuperDARN radar generally show strong backscatter (> 30 dB) from the region of auroral emissions (not shown), which implies the presence of decameter-scale irregularities in the aurora. HF backscatter from visible aurora and precipitation regions have been extensively documented [*Chisham et al.*, 2007, and references therein]. However, since GNSS amplitude scintillation remains low even in the presence of intense auroral emissions and strong HF backscatter, this suggests that in our event the irregularities detected by SuperDARN are not strong enough to cause amplitude scintillation. This was also found in a previous study [*van der Meeren et al.*, 2014] (note, however, that some correspondence between HF backscatter and 250 MHz amplitude scintillation has been found [*Milan et al.*, 2005]).

### 4.3. Spatial Variability of Scintillation Between Two Sites

Another salient point of our observations is the spatial variability of the scintillation. There is a remarkable difference between Longyearbyen and Ny-Ålesund, which are located only ~120 km apart. During Event 2, Longyearbyen (the southernmost of the two locations) reports severe scintillation ($\sigma_\phi > 0.8$) on 6 of 16 visible satellites (Figure 5m). At the same time in Ny-Ålesund (Figure 6m), there is no or only weak scintillation ($\sigma_\phi < 0.3$). The difference between the two locations suggests that intense aurora in the line of sight between the receiver and the satellite is required to produce severe scintillation, as suggested by *Hosokawa et al.* [2014]. Furthermore, it shows that scintillation events and severe ionospheric irregularities can be highly localized and demonstrates the need for a dense network of receivers to properly capture the spatial variability of irregularities. Such localization differences are easily averaged out in statistical studies, where binning may take place on the order of 100–1000km [e.g., *Spogli et al.*, 2009; *Tiwari et al.*, 2012].

### 4.4. Comparison of GNSS Constellations

The arrangement of the GNSS satellites during this event presents a unique opportunity to compare directly scintillation levels across GPS, GLONASS, and Galileo. Figures 5 and 6 show high-resolution 1s scintillation data from all three constellations. As expected due to all constellations using similar frequencies, GPS, GLONASS, and Galileo are similarly affected. Galileo 19 show a maximum phase scintillation of $\sigma_\phi \sim 1.2$, GLONASS 01 show $\sigma_\phi \sim 1.0$, and GPS 09 experience the most severe scintillation at $\sigma_\phi \sim 1.7$. The differences in scintillation levels may be due to (for example) the observational geometry or small-scale differences in line-of-sight emissions, but we cannot make any further conclusions on this based on our data set. It is, however, clear that all constellations experience severe scintillation in a highly localized region of the sky where there are intense auroral emissions causing strong ionospheric irregularities at a wide range of spatial scales.

The level of amplitude scintillation is low across all constellations. There are some very slight enhancements at times seen in Figures 5 and 6, but $S_4$ is always less than 0.2, even in the presence of the intense auroral emissions. The low amplitude scintillation level is in accordance with previous statistical and case studies [*Spogli et al.*, 2009; *Hosokawa et al.*, 2014]. Furthermore, there was no amplitude scintillation ($S_4 < 0.2$) in relation to the patches (data not shown).

## 5. Conclusions

This study has demonstrated where scintillation-producing irregularities may occur in the nighttime polar ionosphere when patches enter the auroral oval during a substorm. The main findings of this study can be summarized as follows:

1. During substorm expansion, severe phase scintillation ($\sigma_\phi > 1$) is observed following the intense poleward edge of the auroral oval as it expands poleward.
2. No phase scintillation ($\sigma_\phi < 0.2$) is observed in relation to two relatively low-density patches exiting the polar cap into the auroral region.
3. No phase scintillation ($\sigma_\phi < 0.2$) is observed in relation to the same patches after having been colocated with intense auroral emissions.
4. Signals may experience strong scintillation when they intersect auroral emissions.
5. In our events the irregularities are most probably driven by intense auroral precipitation. Further studies are needed to ascertain whether patches (blobs) in the region of auroral emissions contribute to the severe scintillation levels.
6. The combination of HF backscatter and low amplitude scintillation suggests that the decameter-scale irregularities causing backscatter are not strong enough to cause amplitude scintillation.
7. Two receivers located ~120 km apart report highly different scintillation impacts, with almost half of the signals scintillating heavily in one receiver and none in the other.
8. GPS, GLONASS, and Galileo were similarly affected by severe phase scintillation in relation to intense line-of-sight auroral emissions in a highly localized region of the sky.

**Acknowledgments**
The University of Oslo ASI data are available at http://tid.uio.no/plasma/aurora. The IMF, solar wind, and *AE* data were provided by the NASA OMNIWeb service (http://omniweb.gsfc.nasa.gov). The Bjørnøya (BJN) magnetometer data were provided by Tromsø Geophysical Observatory, UiT The Arctic University of Norway, and are available from http://flux.phys.uit.no/ascii. The convection data are retrieved from Virginia Tech servers using the DaViTpy software package. GPS data from Hornsund were provided by Marcin Grzesiak and Mariusz Pożoga the Space Research Centre of Polish Academy of Sciences. The GNSS data can be made available upon request from the author. This study was supported by the Research Council of Norway under contracts 212014, 223252, and 230935. We wish to thank Yaqi Jin and Jøran Moen at the University of Oslo for helpful discussions.